\documentclass[a4paper,fleqn,usenatbib,useAMS]{mnras}

\usepackage{graphicx}	% Including figure files
\usepackage{amsmath}	% Advanced maths commands
\usepackage{amssymb}	% Extra maths symbols
\usepackage{multicol}        % Multi-column entries in tables
\usepackage{pdflscape}	% Landscape pages
\usepackage{gensymb}
\usepackage{tikz}
\usetikzlibrary{arrows}

\renewcommand{\eqref}[1]{Eq.~(\ref{#1})}

\newcommand{\bea}{\begin{eqnarray}}
\newcommand{\eea}{\end{eqnarray}}
\newcommand{\beq}{\begin{equation}}
\newcommand{\eeq}{\end{equation}}

\newcommand{\tA}{\tau_{\rm A}}
\newcommand{\vA}{v_{\rm A}}

\newcommand{\tnl}{\tau_{\rm nl}}

\newcommand{\vskipfig}{\vskip-0.25cm}
\newcommand{\dd}{\partial}
\newcommand{\vB}{\mathbf{B}}

\newcommand{\vu}{\mathbf{u}}
\newcommand{\vb}{\mathbf{b}}

\newcommand{\vz}{\mathbf{z}}

\newcommand{\vup}{\vu_\perp}
\newcommand{\vbp}{\vb_\perp}
\newcommand{\vzp}{\vz_\perp}

\newcommand{\dzp}{\delta z_\perp}

\newcommand{\lpar}{l_\parallel}
\newcommand{\Lpar}{L_\parallel}

\newcommand{\hlpar}{\hat{l}_\parallel}
\newcommand{\hlam}{\hat{\lambda}}

\newcommand{\gfkr}{\gamma_{\rm FKR}}
\newcommand{\gcoppi}{\gamma_{\rm Coppi}}

\newcommand{\hltr}{\hlam_{\rm tr}}
\newcommand{\SL}{S_{L_\perp}}
\newcommand{\hlfkr}{\hlam_{\rm FKR}}
\newcommand{\truth}{\tau_{\rm Ruth}}

\newcommand{\Nd}{N_{\rm D}}
\newcommand{\hld}{\hlam_{\rm D}}

\newcommand{\gsp}{\gamma_{\rm SP}}
\newcommand{\gmax}{\gamma_{\rm max}}

\newcommand{\hldpm}{\hat{\lambda}_{{\rm D},i-1}}
\newcommand{\hldp}{\hat{\lambda}_{{\rm D},i}}
\newcommand{\hletap}{\hat{\lambda}_{\eta,i}}
\newcommand{\fsd}{f_{\rm D}}
\newcommand{\feta}{f_{\eta}}
\newcommand{\zom}{{\zeta_m^\perp/m}}
\newcommand{\zot}{{\zeta_2^\perp/2}}
\newcommand{\hldd}{\hlam_{{\rm D},1}}

\usepackage[T1]{fontenc}
\usepackage{ae,aecompl}

\usepackage{mathptmx}
\usepackage{txfonts}

\title[Disruption of sheet-like structures in Alfv\'enic turbulence]{Disruption of sheet-like structures in Alfv\'enic turbulence by magnetic reconnection}

\author[A. Mallet et al.]{A. Mallet$^{1}$\thanks{Contact e-mail: \href{mailto:alfred.mallet@unh.edu}{alfred.mallet@unh.edu}}, A. A. Schekochihin$^{2,3}$, B. D. G. Chandran$^{1}$ \\
$^{1}$Space Science Center, University of New Hampshire, Durham, NH 03824, USA \\
$^{2}$Rudolf Peierls Centre for Theoretical Physics, University of Oxford, Oxford OX1 3NP, United Kingdom\\
$^{3}$Merton College, Oxford OX1 4JD, United Kingdom\\
}

\pubyear{2016}

\begin{document}
\label{firstpage}
\pagerange{\pageref{firstpage}--\pageref{lastpage}}
\maketitle
\begin{abstract}
We propose a mechanism whereby the intense, sheet-like structures naturally formed by dynamically aligning Alfv\'enic turbulence are destroyed by magnetic reconnection at a scale $\hld$, larger than the dissipation scale predicted by models of intermittent, dynamically aligning turbulence. The reconnection process proceeds in several stages: first, a linear tearing mode with $N$ magnetic islands grows and saturates, and
then the $X$-points between these islands collapse into secondary current sheets, which then reconnect until the original structure is destroyed. This effectively imposes an upper limit on the anisotropy of the structures within the perpendicular plane, which means that at scale $\hld$ the turbulent dynamics change: at scales larger than $\hld$, the turbulence exhibits scale-dependent dynamic alignment and a spectral index approximately equal to $-3/2$, while at scales smaller than $\hld$, the turbulent structures undergo a succession of disruptions due to reconnection, limiting dynamic alignment, steepening the effective spectral index and changing the final dissipation scale. The scaling of $\hld$ with the Lundquist (magnetic Reynolds) number $\SL$ depends on the order of the statistics being considered, and on the specific model of intermittency; the transition between the two regimes in the energy spectrum is predicted at approximately $\hld \sim \SL^{-0.6}$. The spectral index below $\hld$ is bounded between $-5/3$ and $-2.3$. The final dissipation scale is at $\hlam_{\eta,\infty}\sim \SL^{-3/4}$, the same as the Kolmogorov scale arising in theories of turbulence that do not involve scale-dependent dynamic alignment. 
\end{abstract}
%%%%%%%%%%%%%%%%%%%%%%%%%%%%%%%%%%%%%%%%%%%%%%%%%%
\section{Introduction}
Turbulence is thought to be important in many astrophysical situations, and is also measured directly by spacecraft in the solar wind \citep{bruno2013}. In many situations, the system consists of an ionized plasma threaded by a strong mean magnetic field $\vB_0$. In this case, the Alfv\'enically polarized fluctuations decouple from the compressive modes and satisfy the reduced magnetohydrodynamics (RMHD) equations \citep{strauss1976}, regardless of the collisionality of the plasma \citep{schektome2009}. Written in terms of \citet{elsasser} variables $\vzp^\pm = \vu_\perp \pm \vb_\perp$, where $\vu_\perp$ and $\vb_\perp$ are the velocity and magnetic field (in velocity units) perturbations perpendicular to $\vB_0$, these equations are
\beq
\dd_t \vzp^\pm \mp v_{\rm A} \dd_z \vzp^\pm + \vzp^\mp \cdot \nabla_\perp \vzp^\pm = -\nabla_\perp p,
\label{eq:RMHD}
\eeq
where the pressure $p$ is obtained from the solenoidality condition $\nabla_\perp\cdot \vz_\perp^\pm = 0$, the Alfv\'en speed is $v_{\rm A} = |\vB_0|$, and $\vB_0$ is in the $z$ direction.

The turbulent system described by Eqs. (\ref{eq:RMHD}) has several interesting characteristics. First, it is anisotropic with respect to the direction of the local magnetic field, as attested by numerical simulations \citep{shebalin83,oughton04,chenmallet,beresnyakanis,mallet3d} and solar-wind measurements \citep{horanis,podestaaniso,wicks10,chenmallet,chen2016}. This anisotropy can be understood in terms of the critical-balance conjecture \citep{gs95,gs97}, whereby the linear (Alfv\'en) time $\tA \sim \lpar / \vA$ ($\lpar$ being the fluctuations' coherence length along the magnetic field line) and nonlinear time $\tnl$ should be similar to each other at all scales, $\tA \sim \tnl$. 

Secondly, it has been noticed that at least in numerical simulations, there is a tendency for the different fields ($\vzp^\pm, \vup, \vbp$) to \emph{align} with one another to within a small, scale-dependent angle $\theta$ \citep{boldyrev,mcatbolalign,bl06}. In the nonlinear term in Eqs. (\ref{eq:RMHD}), only $\vzp^\pm$ with a gradient in the direction of $\vzp^\mp$ gives rise to a nonzero contribution. Combined with the solenoidality of the RMHD fields, this implies that the alignment causes the nonlinearity to be noticeably suppressed. One can take this into account by defining the nonlinear time as follows:
\beq
\tnl^\pm \doteq \frac{\lambda}{\delta z^\mp_\perp \sin\theta}.\label{eq:tnl}
\eeq
If $\sin\theta$ is scale-dependent, it may affect how the fluctuation amplitudes $\dzp^\pm$ scale with the perpendicular scale $\lambda$. 
One can link the alignment effect to local anisotropy of the turbulent structures within the perpendicular plane. The aspect ratio of a sheet is related to the alignment angle between $\delta \mathbf{u}_\perp$  and $\delta\mathbf{b}_\perp$ fluctuations \citep[see ][]{boldyrev} or between $\delta z^+_\perp$ and $\delta z^-_\perp$ fluctuations \citep[see ][]{Chandran14} via%\footnote{The argument that we use here to motivate the link between alignment and anisotropy within the perpendicular plane only really makes sense in a balanced system where $\dzp^+\sim\dzp^-$. In the imbalanced case, a similar result can be obtained by considering the mutual shearing of Elsasser structures \citep[for details, see the appendix of][]{Chandran14}.}. Field lines wander a distance $\xi \sim \lpar {\dzp^\pm}/{\vA}$ in the perpendicular direction within a structure, and, since $\lpar$ is the coherence length in the parallel direction, the structure must remain coherent in the direction of the vector fluctuations up to at least a distance $\xi$. However, since the fields are misaligned by up to an angle $\theta$, the aspect ratio of this structure in the perpendicular plane is
\beq
\frac{\lambda}{\xi} \sim \sin\theta,
\eeq
where $\xi$ is the coherence length of the structure in the direction of the vector fluctuations (henceforth the "fluctuation-direction scale"), and $\lambda$ is the coherence length of the structure in the direction perpendicular to this and also perpendicular to the parallel direction along the magnetic field (which we therefore call the "perpendicular scale"). This 3D anisotropy has been measured in numerical simulations \citep{mallet3d,verdini2015} and in the solar wind \citep{chen3d} (although in the latter case, it has not as yet been definitively pronounced scale-dependent).

The third key feature of Alfv\'enic turbulence, seen in both numerical simulations and in the solar wind, is its high degree of intermittency. Two related models of this intermittency that take into account critical balance and dynamic alignment (\citealt{ms16,Chandran14}, reviewed in Section \ref{sec:turbphen}) both show that, at each scale, higher-amplitude fluctuations are systematically more aligned and, therefore, more anisotropic in the perpendicular plane. Anticorrelation of alignment angle and amplitude has been confirmed in numerical simulations \citep{rcb,mallet3d}. 

Models of turbulence that incorporate dynamic alignment tend to predict perpendicular spectral indices close to $-3/2$ \citep{boldyrev,Chandran14,ms16}, while the original "GS95" \citep{gs95} model, which does not include dynamic alignment, predicts a $-5/3$ spectral index. 
Surprisingly, which of these two classes of models is correct has still not been settled numerically:  while spectral indices measured in extremely high-resolution ($2048^2\times 512$) simulations are very close to $-3/2$ \citep{pmbc,Perez14}, the scaling of the dissipation scale $\lambda_{\eta}$ with Reynolds number ($\mathrm{Re} \doteq L_\perp \delta z / \eta$, where $\eta$ is the resistivity) in simulations with equivalent resolution appears to agree much better with the prediction using the GS95 model, $\lambda_\eta \propto \mathrm{Re}^{-3/4}$ \citep{Beresnyak14}. This suggests that there may be some small scale (perhaps relatively close to, but not smaller than, the dissipative scale predicted by the alignment theories) past which further alignment (or, equivalently, anisotropy within the perpendicular plane) breaks down.

In this paper, we propose a mechanism that causes the turbulent structures to stop aligning and becoming more sheet-like. It appears to be in the nature of the turbulence, at least at large scales, to dynamically generate coherent, large-amplitude, sheet-like structures. It is well known that sheet-like current structures are unstable to tearing modes\footnote{One might also ask whether these sheets could be disrupted by the Kelvin-Helmholtz instability. However, since the vortex stretching terms for the different Elsasser fields have opposite sign \citep{zhdankin16intcy}, in general, there will be more ``current sheets" than ``shear layers". In such sheets, the Kelvin-Helmholtz instability is suppressed \citep{chandrasekharbook}.\label{fn:1}} \citep{fkr,coppi}, and that these modes can eventually disrupt the initial sheet-like structures via magnetic reconnection \citep{loureiro2005,uzdensky2016,tenerani2016}.
%\citep{nuno,bhattacharjee2009,uzdensky2010,loureiro2012,uzdensky2016}. 
This paper attempts to answer the question of whether and at what scale this process occurs for the kind of sheet-like structures that are dynamically formed by the aligning Alfv\'enic turbulence. It has recently been realized that, as current sheets form, they are violently unstable to the tearing instability, and so they never reach the idealized ``Sweet-Parker" reconnection regime \citep{parker1957,sweet1958,nuno,pucci2014,tenerani2016,uzdensky2016} but instead break up into shorter sheets separated by magnetic islands. 
There are several stages in this process: the initial linear growth of the tearing instability, a possible Rutherford stage \citep{rutherford1973} involving secular growth of the magnetic islands, and, finally, collapse of the $X$-points between the islands into short, Sweet-Parker-like sheets, which disrupt the initial structure by magnetic reconnection \citep{loureiro2005}. The characteristic timescales of these processes, discussed in Section \ref{sec:times}, make up the overall time needed to disrupt the sheet, which must be compared with the turbulent cascade time, $\tau_{\rm C}\sim\tnl\sim\tA$, to determine if the disruption occurs. This is done in Section \ref{sec:scales}. We then determine the critical scale below which the sheet-like structures cannot survive, and also determine the number of magnetic islands that the sheets are broken into (see Sections \ref{sec:hld} and \ref{sec:transition}). We also discuss, in Section~\ref{sec:below}, the possible nature of the turbulence below the disruption scale, and show that the disruption process (repeated in a recursive fashion) leads to the \citet{k41} scaling of the final dissipative cutoff, and a steepening of the spectrum below the disruption scale. This can potentially explain the controversy between the results of \cite{Perez14} and \cite{Beresnyak14}, as well as being an interesting physical example of turbulence creating the conditions needed for reconnection. 

\section{Turbulence phenomenology}\label{sec:turbphen}
The intermittency models of \cite{Chandran14} (henceforth CSM15) and \cite{ms16} (henceforth MS17) both envision structures that are characterized by amplitude $\delta z$, and characteristic scales $\lpar$ (parallel scale), $\lambda$ (perpendicular scale) and $\xi$ (fluctuation-direction scale). Here we outline the scalings arising from these models that we will need in this work. Following \cite{ms16}, we introduce normalized variables
\beq
\delta \hat{z} = \frac{\delta z}{\overline{\delta z}}, \quad \hlam = \frac{\lambda}{L_\perp}, \quad \hlpar=\frac{\lpar}{\Lpar}, \quad \hat{\xi} = \frac{\xi}{L_\perp},\label{eq:normalized}
\eeq
where $\overline{\delta z}$ is the outer-scale fluctuation amplitude, and $L_\perp$ and $L_\parallel$ are the perpendicular and parallel outer scales, respectively. The normalized amplitude in both models is given by $\delta \hat{z} \sim \beta^q$, where the non-negative random integer $q$ is a Poisson random variable\footnote{Technically, in the MS17 model, the distribution of $q$ conditional on $\hlam$ is a Poisson mixture, which, however, gives the same scalings for the structure functions as would be obtained with a pure Poisson-distributed $q$.} with the mean $\mu = -\ln\hlam$, and $\beta$ is a dimensionless constant. This form for the distribution of the amplitude may be motivated by modelling the amplitude as decreasing by a fixed factor $\beta$ each time some quantized event (interpreted in CSM15 as a balanced collision) occurs, as the structure sharpens in scale \citep{shewaymire}. The perpendicular scalings are given in both models by
\beq
\langle \delta \hat{z}^m \rangle \sim \hlam^{\zeta_m^\perp},
\eeq
with 
\beq
\zeta_m^\perp = 1-\beta^m,
\eeq
where $\beta$ is fixed via two different strategies in the two models: in MS17, the result is $\beta=1/\sqrt{2}$, while in CSM15, $\beta=0.691$. We will find it useful to define the "effective amplitude" of the structures that dominate the $m$-th order perpendicular structure function:
\beq
\langle \delta \hat{z}^m | \hlam \rangle \equiv (\delta \hat{z}[m])^m \sim \hlam^{1-\beta^m},
\eeq
and so
\beq
\delta \hat{z}[m] \sim \hlam^{(1-\beta^m)/m}.\label{eq:effm}
\eeq
Note that $(1-\beta^m)/m$ is a strictly decreasing function of $m$, and so $m$ is a useful proxy for the amplitude of the structures at a given scale. Three cases in particular will be important. The first is $m\to\infty$, corresponding to the most intense structures with $q=0$, which have amplitudes 
\beq
\delta \hat{z}[\infty] \sim 1,
\eeq
 independent of $\hlam$. Secondly, $m=2$ corresponds to the "r.m.s. amplitude" structures which determine the spectral index (since this is simply related to the scaling of the second-order structure function), and have amplitudes
 \beq
 \delta \hat{z}[2] \sim \hlam^{1/4} \,\text{(MS17)}, \quad \delta \hat{z}[2] \sim \hlam^{0.26} \, \text{(CSM15)}.
 \eeq
Finally, the limit $m\to0$ describes the "bulk" fluctuations with $q=\mu$, whose amplitudes are
\beq
\delta \hat{z}[0] \sim \hlam^{-\ln\beta}.
\eeq
In both models, the fluctuation-direction scale $\hat{\xi}$ is given by
\beq
\hat{\xi} \sim \hlam^{\alpha} \delta \hat{z},\label{eq:xi}
\eeq
and the cascade time is
\beq
\tau_{\rm C} \sim \tnl \sim \hlam^{\kappa_1} \delta \hat{z}^{\kappa_2} \frac{L_\perp}{\overline{\delta z}}.\label{eq:tc}
\eeq
In the MS17 model, 
\beq
\alpha = \kappa_1 = 1/2,\quad\kappa_2 = 0,
\eeq
while in the CSM15 model\footnote{CSM15 defined the quantity $\xi$ (or $\xi_\lambda$ in their notation) to be the characteristic distance along the ($\delta \mathbf{z}^+$) fluctuation direction that a weak $\delta \mathbf{z}^-$ fluctuation would propagate within an intense $\delta \mathbf{z}^+$ sheet before exiting that sheet. CSM15 also showed (see, e.g., their Section 2.6) that two locations within a $\delta \mathbf{z}^+$ sheet that are separated along the fluctuation direction by a distance $\sim \xi_\lambda$ cascade in different and uncorrelated ways. As a sheet-like $\delta \mathbf{z}^+$ structure cascades to smaller scales, its characteristic dimension along the fluctuation direction in the CSM15 model thus becomes $\sim\xi_\lambda$.}, 
\beq
\alpha=1+\ln\beta,\quad \kappa_1 = (1+\ln\beta)^2,\quad\kappa_2 = 1+\ln\beta.
\eeq

 %\footnote{We have glossed over the details of how and why this is done in detail in the models; please refer to \cite{ms16}.} 

Both models envision structures that are sheet-like in the perpendicular plane, with length $\xi$ and width $\lambda$, satisfying $\xi \gg \lambda$. Note that taking $m=2$ in the MS17 model recovers all the scalings of the original dynamic-alignment model due to \cite{boldyrev} (which we will henceforth refer to as B06), but via a different derivation, and positing alignment between Elsasser fields, rather than between velocity and magnetic field. We will assume that the magnetic field varies by $\delta B \sim \delta z$ across the sheet, and further assume that any velocity fluctuation $\delta u\lesssim \delta B$ across the sheet\footnote{The "$\lesssim$" is important because for $\delta u > \delta B$, the Kelvin-Helmholtz mode dominates over the tearing mode. However, we neglect this situation for the reasons given in Footnote \ref{fn:1}.} does not significantly alter the scalings of the tearing instability or its saturation. To determine whether and how structures of a particular amplitude are disrupted faster than they cascade, we must take a detailed look at the different timescales involved in the disruption process.

\section{Timescales}\label{sec:times}
The process whereby a sheet of length $\xi$ and width $\lambda$, with a magnetic field jump $\delta B \sim \delta z$, can be destroyed by reconnection occurs in several stages, which we will now briefly review, following \citet{uzdensky2016}. First, there is exponential growth of the linear tearing mode until the width of the island(s) is approximately the width of the inner layer where resistivity is important, 
\beq
w \sim \delta_{\rm in} \sim \left[ \gamma (k \delta z)^{-2} \lambda^2 \eta \right]^{1/4},\label{eq:deltain}
\eeq
where $\gamma$ is the linear growth rate of the tearing mode and $k\sim N / \xi$ is its wavenumber ($N$ is the number of islands). Secondly, there may be secular ``Rutherford" growth of the islands until $w \sim 1/\Delta'$, where $\Delta'$ is the instability parameter for the tearing mode. Thirdly, the $X$-point(s) that have arisen collapse into thin sheets, which then undergo fast reconnection, leaving behind a set of magnetic islands. We will examine these processes to determine which of them dominates the total time to disrupt the sheet, and thus determine whether this is faster than the cascade time $\tau_{\rm C}$.

\subsection{Linear growth stage}\label{sec:lingrow}
We will assume that a typical sheet-like structure arising in dynamically aligning turbulence is reasonably well modelled by a \citet{harris1962} sheet, so
\beq
\Delta'\lambda = 2\left( \frac{1}{k\lambda} - k\lambda \right).\label{eq:dprime}
\eeq
There is an instability provided that $\Delta'>0$. We are interested in long-wavelength modes, so
\beq
\Delta' \lambda \sim \frac{1}{k\lambda} \sim \frac{\hat{\xi}}{N\hat{\lambda}}.
\eeq
%where $N$ is the number of magnetic islands formed by the linear instability, $N \sim k\xi$. 
There are two possible situations: (i) $\Delta'\delta_{\rm in} \ll 1$, "FKR" modes \citep{fkr} with 
\begin{align}
\gfkr &\sim \Delta'^{4/5} k^{2/5}\delta z^{2/5} \lambda^{-2/5} \eta^{3/5}\nonumber\\
&\sim N^{-2/5} \left(\frac{{\xi}}{\lambda}\right)^{2/5} S_\lambda^{-3/5} \frac{\delta z}{\lambda},\label{eq:fkr}
\end{align}
and (ii) $\Delta'\delta_{\rm in} \sim 1$, "Coppi" modes \citep{coppi} with 
\begin{align}
\gcoppi &\sim k^{2/3} \delta z^{2/3} \lambda^{-2/3} \eta^{1/3}\nonumber\\
&\sim N^{2/3} \left(\frac{{\xi}}{\lambda}\right)^{-2/3} S_\lambda^{-1/3} \frac{\delta z}{\lambda},
\end{align}
where the Lundquist number\footnote{Note that $S_\lambda$ is just what in turbulence theory one would usually call the local magnetic Reynolds number at scale $\lambda$.} is $S_\lambda \doteq \lambda \delta z / \eta$. Since these two modes have opposite dependence on $k$, the maximum growth rate can be found at the wavenumber where $\gfkr \sim \gcoppi$, giving 
\beq
k_{\rm max} \lambda \sim S_{\lambda}^{-1/4}, \quad \gamma_{\rm max} \sim S_\lambda^{-1/2} \frac{\delta z}{\lambda}.\label{eq:kmaxcoppi}
\eeq
However, this "transitional mode" is only accessible if it actually fits into the sheet, i.e., if 
\beq
k_{\rm max} \xi\sim N_{\rm max} \sim \frac{\hat{\xi}}{\hlam} S_\lambda^{-1/4} > 1.\label{eq:nmax}
\eeq

The maximum growth rate for a particular structure is thus either $\gamma_{\rm max}$ given by Eq.~(\ref{eq:kmaxcoppi}), if $k_{\rm max}$ fits into the structure, or $\gfkr$ given by Eq.~(\ref{eq:fkr}) with $N=1$, the longest-wavelength FKR mode, otherwise.
\subsection{Rutherford growth stage}\label{sec:ruth}
The linear growth stage ends when the width of the islands $w\sim \delta_{\rm in}$, given by Eq.~(\ref{eq:deltain}). If $\Delta'w \ll 1$, there will be secular "Rutherford" growth \citep{rutherford1973} until $\Delta'w \sim 1$,
\beq
w \sim \eta \Delta' t.%\dot{w} \simeq \eta \Delta'.
\eeq
If present, this stage lasts for a time
\beq
\tau_{\rm Ruth} \sim \frac{1}{\eta \Delta'^2} \sim N^2 \left(\frac{\hlam}{\hat{\xi}}\right)^2 S_\lambda \frac{\lambda}{\delta z}.
\eeq
Note that $\truth$ increases with $N$, so, if the maximum growth rate is attained for the $N=1$ FKR mode, this mode will also exit the Rutherford stage and saturate first. In the FKR limit,
$\Delta'\delta_{\rm in} \ll 1$,
and so there is a well-defined Rutherford stage. For the Coppi modes, $\Delta'\delta_{\rm in} \sim 1$, and so there is no Rutherford stage.

\subsection{Collapse/reconnection stage}\label{sec:coll}
At the end of the Rutherford stage (or immediately after the linear stage in the case of Coppi modes), the $X$-point(s) formed by the tearing mode collapse into thin secondary sheets, each of length $\sim \xi/N$, and reconnect the flux in the original structure. This collapse, studied by \cite{loureiro2005}, results in exponential, Sweet-Parker-like growth of the reconnected flux on a timescale that, written in terms of our variables, is
\beq
\gsp \sim S_{\lambda}^{-1/2} \frac{\delta\hat{z}}{\lambda} \sim \gmax,\label{eq:gcoll}
\eeq
and so the rate of the collapse is always greater than or equal to the growth rate of the initial linear instability. Following \cite{uzdensky2016}, we therefore do not need to consider the time associated with this stage in our determination of the disruption time of the original structure.

For high enough $\SL$, the collapse rate becomes independent of $S_{\lambda}$ because the Lundquist number associated with the secondary sheets becomes larger than $S_{\rm c}\sim10^4$, the critical Lundquist number required to trigger the onset of plasmoid-dominated fast reconnection \citep{nuno,samtaney2009,bhattacharjee2009,uzdensky2010,loureiro2012}. The critical $\SL$ necessary to access the plasmoid-dominated regime will be determined in Section \ref{sec:plasmoids}.

\subsection{Disruption time}
Based on the above scalings, we can now identify the disruption time as
\beq
\tau_{\rm D} \sim \begin{cases}
\mathrm{max}[1/\gfkr,\truth] & \text{if $\hlam > \hltr$},\\
1/\gmax & \text{if $\hlam \leq \hltr$}.\label{eq:td}
\end{cases}
\eeq
This is just restating the key result of \cite{uzdensky2016}, which will allow us to compare $\tau_{\rm D}$ with the cascade time $\tau_{\rm C}$. The transition scale $\hltr$ will be worked out in Section \ref{sec:hltr}. 

\subsection{Resistive time}
If the structures are not able to be disrupted by reconnection, they can simply decay resistively on a timescale
\beq
\tau_\eta \sim \frac{\lambda^2}{\eta}.\label{eq:teta}
\eeq
The interesting question that we will answer in this paper is whether and under what circumstances this basic dissipation mechanism is superceded by tearing and the onset of reconnection.

\section{Critical scales}\label{sec:scales}
In this section, we will calculate the critical scales that partition the domain defined by $\hlam$ and $m$ [see Eq.~(\ref{eq:effm})] into regions where the structures are and are not disrupted by the onset of reconnection. To do this, we need to compare the timescales identified in section \ref{sec:times} to the cascade time $\tau_{\rm C}$ [Eq.~(\ref{eq:tc})].

\subsection{Resistive scale}\label{sec:hleta}
First, we deal with the dissipative scale for the turbulence in the absence of any disruption by reconnection. Using Eqs.~(\ref{eq:tc}) and (\ref{eq:teta}), we evaluate
\beq
\frac{\tau_{\rm C}}{\tau_\eta} \sim \frac{\hlam^{\kappa_1}\delta \hat{z}^{\kappa_2} {L_\perp}\eta}{\lambda^2 {\overline{\delta z}}}\sim \hlam^{\kappa_1 -2}\delta\hat{z}^{\kappa_2} \SL^{-1}.
\eeq
%where $\SL\doteq L_\perp \overline{\delta z}/\eta$ is the global (outer scale) Lundquist/Reynolds number. 
We consider fluctuations $\delta\hat{z}[m]$ that are important for the $m$th-order structure function, using Eq. (\ref{eq:effm}), to obtain
\beq
\frac{\tau_{\rm C}}{\tau_\eta} \sim \SL^{-1}\hlam^{\kappa_1-2+\kappa_2\zeta_m^\perp/{m}}.
\eeq
Therefore, the resistive scale for these $m$th-order structures is
\beq
\hlam_\eta \sim \SL^{-(2-\kappa_1 - \kappa_2{\zeta_m^\perp}/{m})^{-1}}.
\eeq
In the MS17 model, since $\kappa_1 = 1/2$ and $\kappa_2=0$,
\beq
\hlam_\eta^{\rm MS} \sim \SL^{-2/3},\label{eq:hletams}
\eeq
independent of $m$. This is the standard estimate for the dissipation scale (the analogue of the Kolmogorov scale) in the original dynamic-alignment model of B06 %of \cite{boldyrev} 
(see, e.g., \citealt{pmbc} for an explicit derivation of this scaling). In the CSM15 model, 
\beq
\hlam_\eta^{\rm CSM} \sim \SL^{-\left(1.60-0.63{\zeta_m^\perp}/{m}\right)^{-1}},\label{eq:hletacsm}
\eeq
so the low-order, lower-amplitude fluctuations dissipate at smaller scales than the high-order, higher-amplitude fluctuations.

\subsection{Boundary between FKR and transitional modes}\label{sec:hltr}
The boundary between the two different regimes for the linear tearing stage is given by Eq.~(\ref{eq:nmax}), $S_\lambda^{-1/4}\hat{\xi}/\hlam \sim 1$. Using Eq.~(\ref{eq:xi}) and replacing $\delta \hat{z}$ with the typical amplitude of an $m$th-order structure given by Eq.~(\ref{eq:effm}), we see that the transitional mode (\ref{eq:kmaxcoppi}) may only occur when
%\beq
%\hlam < \hltr \sim \left[ \SL^{1/4}N\right]^{\left(\alpha - \frac{5}{4}+\frac{3}{4}\frac{\zeta_m^\perp}{m}\right)^{-1}}.
%\eeq
\beq
\hlam < \hltr \sim \SL^{\frac{1}{4}\left(\alpha - \frac{5}{4}+\frac{3}{4}\frac{\zeta_m^\perp}{m}\right)^{-1}}.\label{eq:hltr}
\eeq
In the MS17 model, 
\beq
\hltr^{\rm MS} \sim \SL^{-\frac{1}{3}(1-\zeta_m^\perp/m)^{-1}},\label{eq:hltrms}
\eeq
while in the CSM15 model,
\beq
%\hltr^{\rm CSM} \sim \SL^{\frac{1}{4}(-0.62+3\zeta_m^\perp/4m)^{-1}}.\label{eq:hltrcsm}
\hltr^{\rm CSM} \sim \SL^{-0.40(1 - 1.21\zeta^\perp_m/m)^{-1}}.\label{eq:hltrcsm}
\eeq
At scales $\hlam>\hltr$, the FKR mode with $N=1$ is the most unstable linear mode, while at $\hlam<\hltr$, the transitional mode (\ref{eq:kmaxcoppi}) is the fastest.
\begin{figure*}
\centering
\vskipfig
\begin{tabular}{cc}
\includegraphics[width=0.485\textwidth]{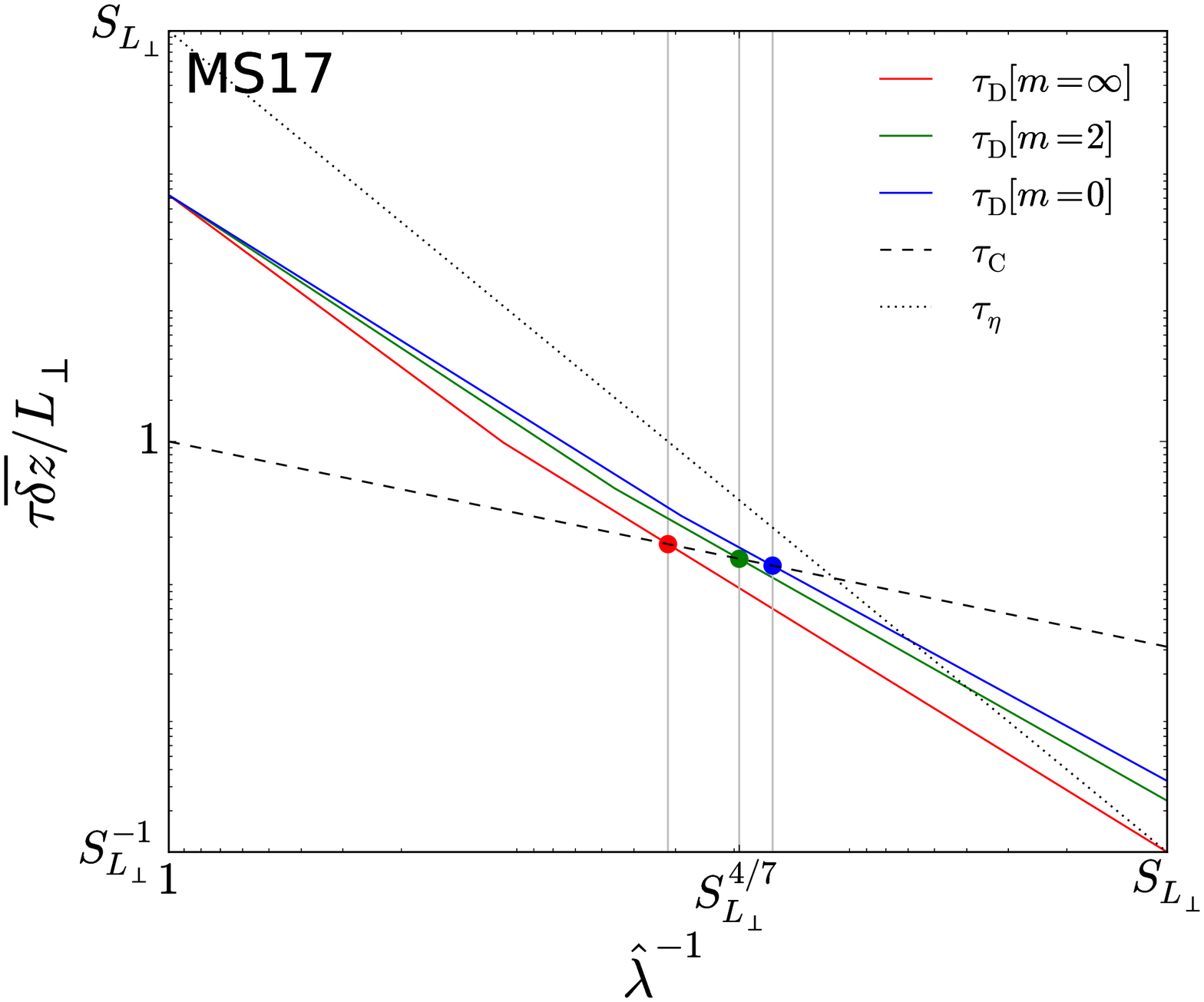} &
\includegraphics[width=0.485\textwidth]{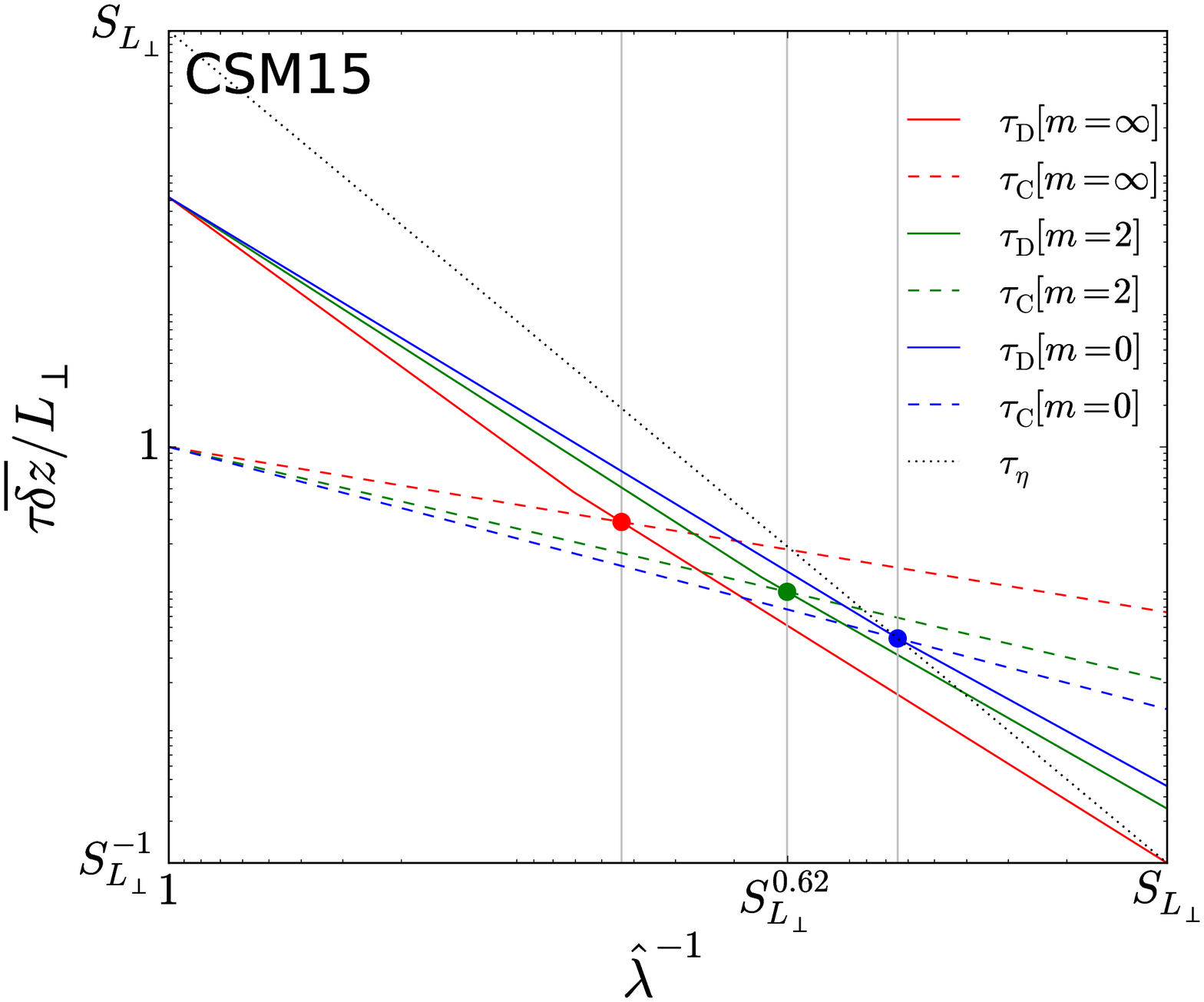}
\end{tabular}
\vskipfig
\caption{Comparison of the cascade timescale $\tau_{\rm C}$, Eq.~(\ref{eq:tc}) (dashed lines), the disruption timescale $\tau_{\rm D}$, Eq.~(\ref{eq:td}) (solid, coloured lines) and the resistive timescale $\tau_{\eta}$, Eq.~(\ref{eq:teta}) (dotted lines), for the MS17 model (left panel) and the CSM15 panel (right panel). Where these timescales depend on the order $m$ of the fluctuations, three values of $m$ are plotted: $m=\infty$ (in red), $m=2$ (in green) and $m=0$ (in blue). In the CSM15 model, not only $\tau_{\rm D}$ but also the cascade time varies with $m$, and so there are three curves for $\tau_{\rm C}$. In the MS17 model, $\tau_{\rm C}$ does not depend on $m$, and so there is a single curve. The point at which the disruption process becomes faster than the turbulent cascade is marked with a circle for each $m$, and a gray vertical line marks the corresponding scale $\hld[m]$, given by Eqs.~(\ref{eq:ndms}) for MS17 and (\ref{eq:ndcsm}) for CSM15. \label{fig:timescales}}
\vskipfig
\end{figure*}

\subsection{Linear FKR critical scale}\label{sec:hlfkr}
To determine whether the $N=1$ FKR mode grows fast enough to disrupt the structures, we first calculate, using Eqs. (\ref{eq:effm}), (\ref{eq:xi}), (\ref{eq:tc}) and (\ref{eq:fkr}),
\begin{align}
\gfkr[N=1] \tau_{\rm C} &\sim \left(\frac{\hat{\xi}}{\hlam}\right)^{2/5} S_\lambda^{-3/5} \frac{\delta \hat{z}}{\hlam} \hlam^{\kappa_1} \delta \hat{z}^{\kappa_2}\nonumber \\
&\sim  \SL^{-3/5}\hlam^{\frac{2\alpha}{5}-2+\kappa_1}\delta\hat{z}^{\frac{4}{5}+\kappa_2}\nonumber\\
&\sim  \SL^{-3/5}\hlam^{\frac{2\alpha}{5}-2+\kappa_1 + \frac{\zeta_m^\perp}{m}\left(\frac{4}{5}+\kappa_2\right)}.
\end{align}
The sheet will not be disrupted unless $\gfkr[N=1] \tau_{\rm C} > 1$. This is equivalent to 
\beq
\hlam<\hlfkr \sim \SL^{\frac{3}{5}\left[\frac{2\alpha}{5}-2+\kappa_1 + \frac{\zeta_m^\perp}{m}\left(\frac{4}{5}+\kappa_2\right)\right]^{-1}}.
\eeq
For the MS17 model,
\beq
\hlfkr^{\rm MS} \sim \SL^{-\frac{6}{13}\left(1-\frac{8\zeta_m^\perp}{13m}\right)^{-1}},\label{eq:hlfkrms}
\eeq
while for the CSM15 model,
\beq
%\hlfkr^{\rm CSM} \sim \SL^{-\frac{3}{5}\left(1.35-\frac{1.43\zeta_m^\perp}{m}\right)^{-1}}.\label{eq:hlfkrcsm}
\hlfkr^{\rm CSM} \sim \SL^{-0.44(1-{1.06\zeta_m^\perp}/{m})^{-1}}.\label{eq:hlfkrcsm}
\eeq
Comparing these scalings with Eq.~(\ref{eq:hltr}), we see that the scale $\hlfkr$ is smaller than the corresponding $\hltr$ for all $m$, and so there are no FKR modes that grow fast enough to disrupt the structures. Therefore, we do not need to consider the duration of the Rutherford stage (see Section \ref{sec:ruth}) to determine whether disruption occurs.

\subsection{Disruption scale}\label{sec:hld}
For a given $m$, at $\hlam \leq \hltr$, with the latter scale given by Eq.~(\ref{eq:hltr}), the disruption time is, therefore, 
\beq
\tau_{\rm D} \sim 1/\gamma_{\rm max},\label{eq:td2}
\eeq
and we must calculate
\begin{align}
\gmax\tau_{\rm C} &\sim S_{\lambda}^{-1/2} \frac{\delta \hat{z}}{\hlam}\hlam^{\kappa_1}\delta \hat{z}^{\kappa_2}\nonumber \\
&\sim \hlam^{-3/2+\kappa_1 + (1/2+\kappa_2){\zeta_m^\perp}/{m}}\SL^{-1/2},
\end{align}
where we have used Eq. (\ref{eq:kmaxcoppi}) for $\gmax$ and Eqs.~(\ref{eq:effm}) and (\ref{eq:tc}) to express $\tau_{\rm C}$ and $\delta \hat{z}$ in terms of $\hlam$ and $m$. The sheet will be disrupted if $\gmax\tau_{\rm C} > 1$. This happens for
\beq
\hlam < \hld \sim \SL^{-\frac{1}{2}\left[\frac{3}{2}-\kappa_1 - \left(\frac{1}{2}+\kappa_2\right)\frac{\zeta_m^\perp}{m}\right]^{-1}}.
\eeq
The corresponding number of islands, from Eq.~(\ref{eq:nmax}), is
\begin{align}
\Nd &\sim S_{\lambda_{\rm D}}^{-1/4} \frac{\hat{\xi}_{\rm D}}{\hld},\nonumber\\
&\sim \hld^{\alpha - 5/4 +{3\zeta_m^\perp}/{4m}}\SL^{-1/4},\nonumber\\
&\sim \SL^{\left\{\frac{1}{2}\left(\frac{5}{4}-\alpha -\frac{3\zeta_m^\perp}{4m}\right)\left[\frac{3}{2}-\kappa_1 - \left(\frac{1}{2}+\kappa_2\right)\frac{\zeta_m^\perp}{m}\right]^{-1}-\frac{1}{4}\right\}}.
\end{align}
In the MS17 model, these scalings become
\beq
\hld^{\rm MS} \sim \SL^{-\frac{1}{2}\left(1-\frac{\zeta_m^\perp}{2m}\right)^{-1}}, \quad \Nd^{\rm MS} \sim \SL^{\frac{1-2\zeta_m^\perp / m}{8-4\zeta_m^\perp/m}},\label{eq:ndms}
\eeq
while in the CSM15 model,
\beq
%\hld^{\rm CSM} \sim \SL^{-0.45\left(1-1.03{\zeta_m^\perp}/{m}\right)^{-1}}, \quad \Nd^{\rm CSM} \sim \SL^{-\frac{1}{4} + \frac{0.309-{3\zeta_m^\perp}/{8m}}{1.10 - 1.13{\zeta_m^\perp}/{m}}}.\label{eq:ndcsm}
\hld^{\rm CSM} \sim \SL^{-0.45\left(1-1.03{\zeta_m^\perp}/{m}\right)^{-1}}, \quad \Nd^{\rm CSM} \sim \SL^{\frac{1-{2.71\zeta_m^\perp}/{m}}{32.3 - 32.1{\zeta_m^\perp}/{m}}}.\label{eq:ndcsm}
\eeq
Note that $\hld < \hltr$, as expected, since no FKR modes grow fast enough to disrupt the sheets (see Section \ref{sec:hlfkr}). These scalings determine the largest $\hlam$ for which the fastest-growing mode reaches collapse in a time shorter than the cascade time $\tau_{\rm C}$ of the turbulence, and, therefore, the smallest $\hlam$ for which the aligned, sheet-like structures can survive. We will examine some instructive particular cases and the physical consequences of these results in Sections \ref{sec:transition} and \ref{sec:below}.

\subsection{Critical $\SL$ for the plasmoid-dominated regime}\label{sec:plasmoids}
As an interesting aside, we noted in Section 
\ref{sec:coll} that for high enough $\SL$, the reconnection rate $\gamma_{\rm SP}$ becomes independent of $S_{\lambda}$ due to the onset of the plasmoid instability. For this to occur, 
the Lundquist number associated with the secondary sheets must be
\beq
S_{\xi_{\rm D}/\Nd} \sim \hld \delta \hat{z}_{\hld} S_{\hld}^{1/4}\SL\sim \hld^{\frac{5}{4}\left(1+\zom\right)} \SL^{5/4} > S_{\rm c},
\eeq
%the aspect ratio of the secondary sheets must be 
%\beq
%R\sim \frac{\xi}{N\lambda} \sim S_{\lambda}^{1/4} > R_{\rm crit}.
%\eeq
where we used Eq.~(\ref{eq:nmax}) for $\Nd=N_{\rm max}$. Expressing this condition in terms of $\SL$, we obtain in the MS17 model, using Eq.~(\ref{eq:ndms}),
\beq
\SL > S_{\rm c}^{\frac{4}{5}\frac{1-\zeta_m^\perp/2m}{1/2-\zeta_m^\perp/m}}.\label{eq:critsl}
\eeq
%%\beq
%%\SL^{\frac{1+2\zom}{8-4\zom}} > R_{\rm crit},
%%\eeq
%%so that for 
%\beq
%\SL> R_{\rm crit}^{\frac{8-4\zom}{1+2\zom}}.\label{eq:critsl}
%\eeq
In the CSM15 model, we obtain
\beq
\SL > S_{\rm c}^{ \frac{1.47 - 1.51 \zom}{1 - 2.72 \zom}}.
\eeq
For such values of $\SL$, the secondary sheets will break into plasmoids and the reconnection/collapse rate will be given by
\beq
\gamma_{\rm plasmoids} \sim S_{\rm c}^{-1/2} \frac{\delta \hat{z}}{\lambda},
\eeq
instead of Eq.~(\ref{eq:gcoll}), because the secondary sheet will be broken into "critical Sweet-Parker sheets" \citep{uzdensky2010}, each of which will reconnect at this rate. Assuming $S_{\rm c} \sim 10^4$ \citep{nuno,samtaney2009},\footnote{Note that in a turbulent environment, $S_{\rm c}$ may be somewhat lower, possibly by as much as an order of magnitude \citep[see ][]{loureiro2009}.} the critical $\SL$ given by Eq.~(\ref{eq:critsl}) is quite high: for the $m=\infty$ structures in the MS17 model to be plasmoid unstable, $\SL \gtrsim S_{\rm c}^{8/5} \sim 10^{6}$, while for the $m=2$ structures in the MS17 model, $\SL \gtrsim S_{\rm c}^{14/5} \sim 10^{11}$. In the CSM15 model, the $m=\infty$ structures are plasmoid unstable for $\SL\gtrsim S_{\rm c}^{1.47}\sim 10^6$, while the $m=2$ structures are plasmoid unstable for $\SL\gtrsim S_{\rm c}^{3.7}\sim10^{15}$. This suggests that the plasmoid-dominated regime is not accessible in current numerical simulations, as indeed confirmed by the absence of plasmoid-unstable current sheets in the simulations of \cite{zhdankin13}. The critical $\SL$ is much higher than the critical Lundquist number for a Sweet-Parker sheet to be plasmoid unstable because the structures formed by the turbulence do not have a particularly high aspect ratio. The mechanism outlined in this paper does not rely on the secondary sheets being plasmoid unstable: for the disruption to occur, we only need $\tau_{\rm C}/\tau_{\rm D} > 1$, where $\tau_{\rm D}$ is set by the tearing growth rate.

\section{Transition to a new regime of strong Alfv\'enic turbulence}\label{sec:transition}
The comparison of timescales in Section \ref{sec:scales} has allowed us to predict the scale at which the sheet-like structures at each order $m$ are disrupted by the onset of reconnection. While the cascade time $\tau_{\rm C}$ [Eq.~(\ref{eq:tc})] decreases as the cascade progresses to smaller scales, so does the disruption time $\tau_{\rm D}$ [Eq.~(\ref{eq:td})]. Since the nonlinearity in the aligned sheet-like structures is suppressed by a factor equal to their alignment angle (inverse aspect ratio), $\tau_{\rm D}$ decreases faster than the cascade time, and eventually becomes smaller, at the scale $\hld$. This is shown in Figure \ref{fig:timescales} for both the MS17 and CSM15 models, for $m=\infty, 2, 0$ (most intense, r.m.s. amplitude, and most typical structures, respectively). Also shown are the disruption scales $\hld[m]$ beyond which the sheet-like structures of order $m$ cannot survive. 

\begin{figure*}
\vskipfig
\begin{tabular}{cc}
\includegraphics[width=0.48\linewidth]{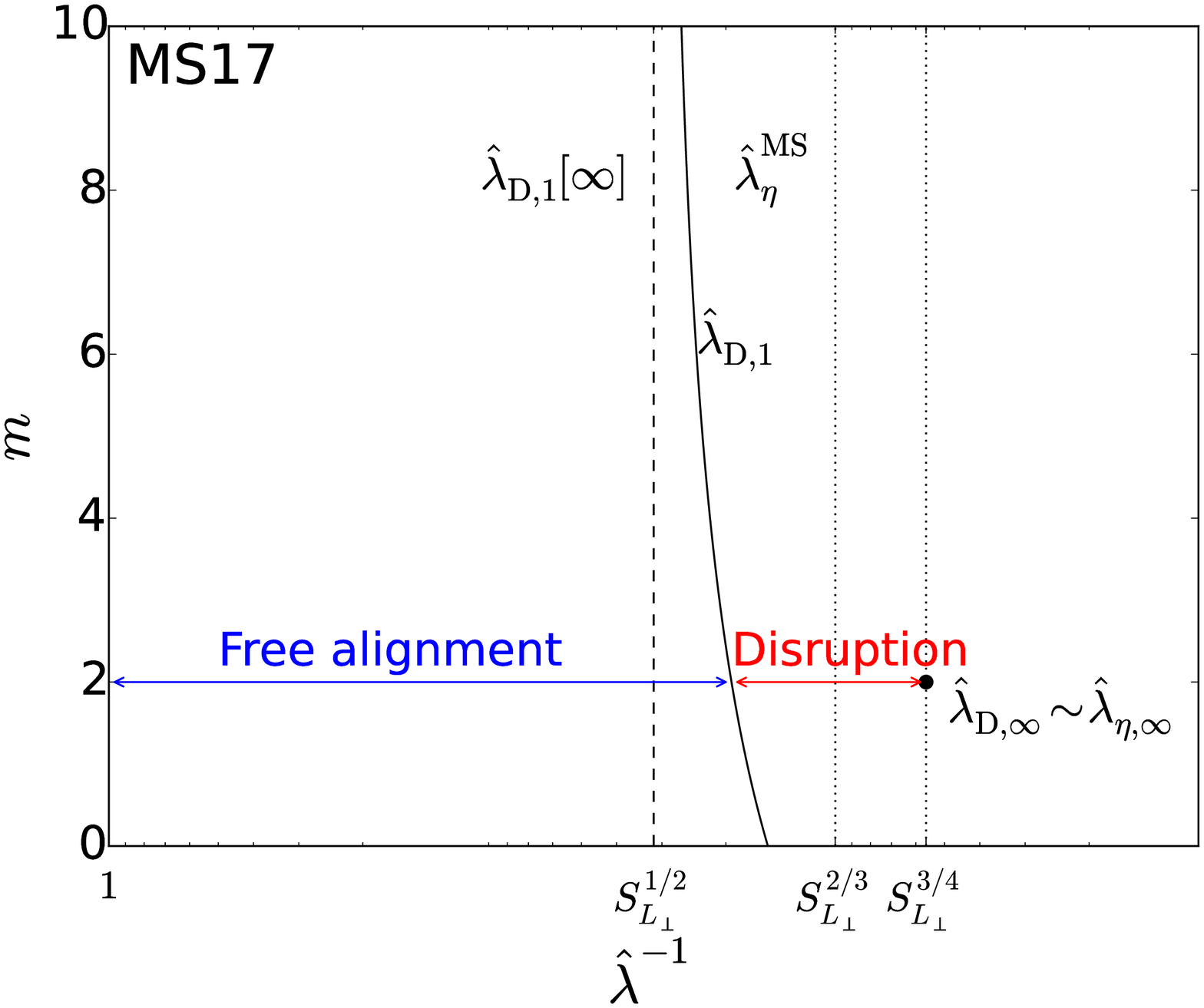}&
\includegraphics[width=0.48\linewidth]{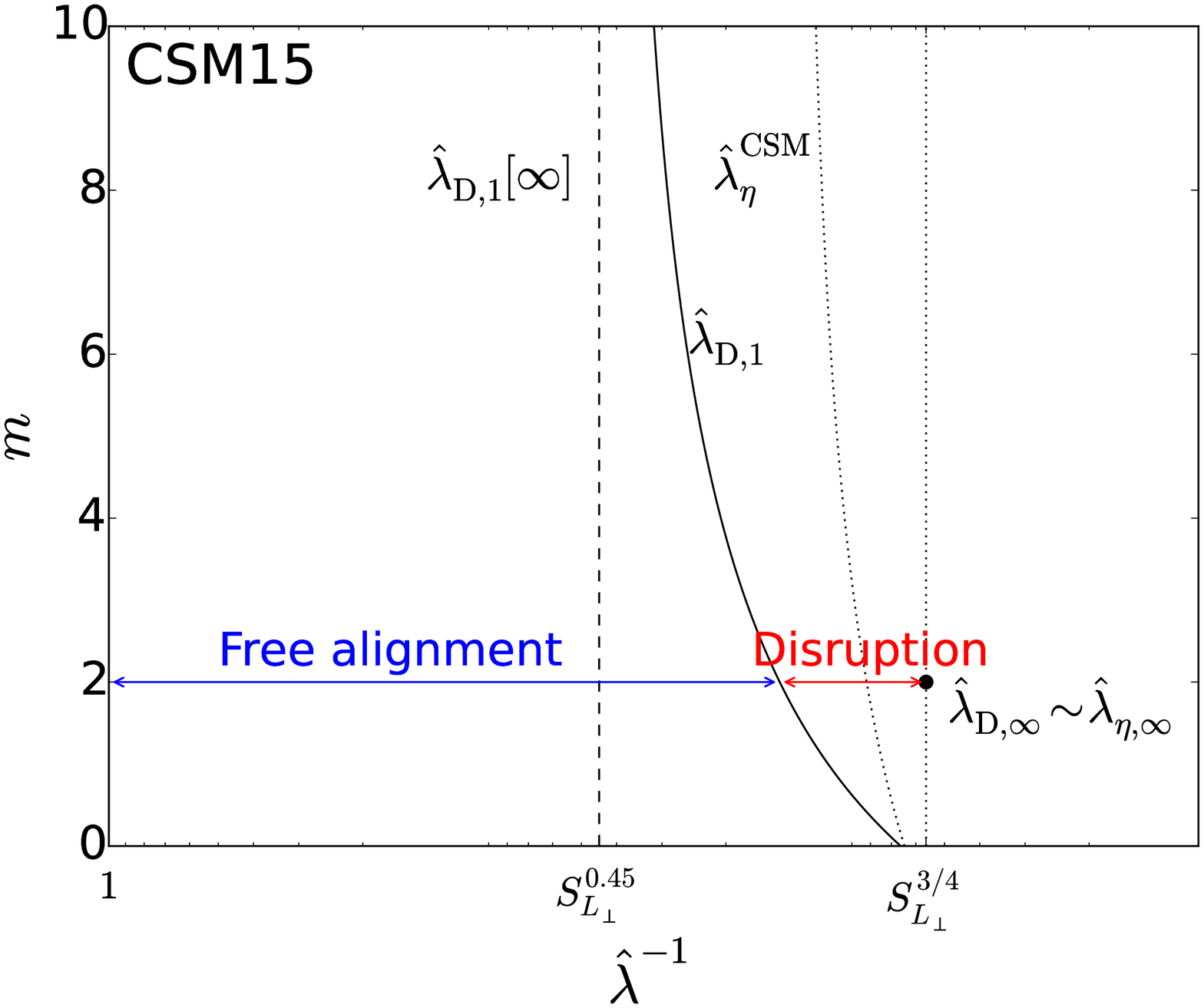}\\
\end{tabular}
\vskipfig
\caption{The solid black line shows the dependence of $\hlam_{{\rm D},1}$ on $m$ in the MS17 model [Eq.~(\ref{eq:ndms}), left panel] and in the CSM15 model [Eq.~(\ref{eq:ndcsm}), right panel]. The dashed lines show $\hlam_{{\rm D},1}[m=\infty]$. The two different subranges for the $m=2$ structures are marked by blue and red arrows. The leftmost (i.e., larger scale) dotted line on each plot shows the expected dissipation scales (\ref{eq:hletams}) and (\ref{eq:hletacsm}) of the Alfv\'enic turbulence without disruption, while the rightmost (smaller scale) dotted line shows the Kolmogorov scale, $\SL^{-3/4}$. The black point shows the value of the final dissipation scale of the disrupted turbulence $\hlam_{{\rm D},\infty} \sim \hlam_{\eta,\infty}$ [Eqs.~(\ref{eq:hldinf}), (\ref{eq:hletainf})], equal to the Kolmogorov scale.\label{fig:cascade}}
\vskipfig
\end{figure*}

The effect that the disruption has on the turbulence is, thus, as follows: for $\hlam < \hld$, the sheet-like structures predicted by the turbulence models that rely on dynamic alignment (e.g., MS17, CSM15, and the original model of %\citealt{boldyrev}
B06) are disrupted by reconnection into several separate islands. The detailed dependence of $\hld$ on $m$ in the MS17 and CSM15 models [Eqs.~(\ref{eq:ndms}) and (\ref{eq:ndcsm}), respectively] is shown in Figure \ref{fig:cascade}, along with the resistive scales corresponding to these models [Eqs.~(\ref{eq:hletams}) and (\ref{eq:hletacsm})]. The disruption scale $\hld$ is an increasing function of $m$.  Roughly speaking, one might expect the behaviour of the $m$th-order structure function to change at $\hld[m]$. In practice, since structures of all orders contribute to all structure functions to differing degrees, the transition will take place over a range of scales between $\hld[\infty]$ and $\hld[0]$. As $m\to\infty$, the disruption scale approaches
\beq
\hld^{\rm MS}[\infty] \sim \SL^{-1/2}, \quad \hld^{\rm CSM}[\infty] \sim \SL^{-0.45},
\eeq
in the MS17 and CSM15 models, respectively. One might expect to see a change in the spectral index (since this is related to the scaling exponent of the second-order structure function) at around
\beq
\hld^{\rm MS}[2] \sim \SL^{-4/7}, \quad \hld^{\rm CSM}[2] \sim \SL^{-0.62}.\label{eq:ldmeq2}
\eeq
For $m=0$ structures, the disruption scale is given by
\beq
\hld^{\rm MS}[0] \sim \SL^{-0.60}, \quad \hld^{\rm CSM}[0] \sim \SL^{0.73}.
\eeq
In the MS17 model, the disruption scale is above the resistive scale for all $m$, $\hld > \hlam_{\eta}$ (see Section \ref{sec:hleta}). In the CSM15 model, $\hld>\hlam_{\eta}$ for all $m>0$, but $\hld[0]$ and $\hlam_{\eta}[0]$ are identical. Thus, in the CSM15 model, $m=0$ structures, which are neither aligned nor sheet-like, cascade to their resistive scale without being disrupted by the onset of reconnection. This explicitly shows that the suppression of the nonlinearity due to dynamic alignment is required for the disruption process to become effective at a larger scale than $\hlam_{\eta}$.

\section{Turbulence below $\hld$}\label{sec:below}
It is natural to ask what happens to the turbulence below the disruption scale $\hld$. We will restrict ourselves to the case of $m=2$ (the r.m.s. amplitude structures) for the following discussion, i.e., we forgo any discussion of intermittency below $\hld$.

We expect the sheet-like structures just above $\hld$ to be broken up into "flux-rope-like" structures (3D versions of plasmoids) just below $\hld$: these are roughly circular in the perpendicular plane, with scale $\hld$,\footnote{Based on the numerical evidence in \cite{loureiro2005}, this does appear to be how the tearing mode saturates at high enough $\Delta'$.} %: they observed a plateau at high $\Delta'$ of $W_{\rm sat} \sim L_\perp$.} 
but extended in the direction parallel to the local magnetic field, due to critical balance. These structures, no longer anisotropic in the perpendicular plane, will break up nonlinearly, serving as the energy-containing "eddies" of a new cascade. 

This implies that the fluctuation amplitude just below $\hld$ should decrease\footnote{This does not mean that there are actually sharp jumps in the structure function. As the cascade progresses to smaller scales, the fraction of the energy contained in disrupted structures increases continuously: the disruption scale is just the scale at which a given structure function is dominated by disrupted structures.}. Indeed, the energy flux just below the scale $\hld$ must be equal to the energy flux just above it, or at any other scale in the inertial range:
\beq
\epsilon \sim \frac{\overline{\delta z}^3}{L_\perp} \sim \frac{\delta z_{1,-}^3}{\lambda_{\rm D}},
\eeq
where $\delta z_{1,-}$ is the amplitude of the new structures. This gives a simple expression for this dynamically adjusted amplitude:
\beq
\delta \hat{z}_{1,-} \sim \hld^{1/3},\label{eq:under}
\eeq
where we have normalized by outer-scale quantities in the usual way (\ref{eq:normalized}). We have assumed here that the reconnection process involved in the $X$-point collapse and formation of flux ropes (plasmoids) can be viewed as mostly transferring energy from one form of magnetic/velocity perturbation at scale~$\lambda_{\rm D}$ (aligned structures) to another form of perturbation at scale~$\lambda_{\rm D}$ (plasmoids, outflows). 
 Moreover, since the cascade timescale $\lambda_{\rm D}/\delta z_{1,-}$  for unaligned structures just below scale~$\lambda_{\rm D}$ is shorter than the disruption timescale, we assume that nonlinear interactions between unaligned structures are the dominant mechanism for transferring fluctuation energy from scale~$\lambda_{\rm D}$ to smaller scales.
 If a constant energy flux across $\hld$ were not a good assumption, the amplitude below $\hld$ would be smaller than in Eq.~(\ref{eq:under}), and the corresponding spectral slope at scales below $\hld$ steeper than will be deduced in Section \ref{sec:spectrum}.

We expect the new structures to behave as they normally would in Alfv\'enic turbulence: to interact, cascade to smaller scales, and dynamically align as the scale decreases. The change compared to the "primary cascade" is that the disruption process effectively resets the perpendicular anisotropy at scale $\hld$, so the aligning structures have smaller aspect ratios than they would have had without the disruption. The amplitude of the $(m=2)$ turbulent structures at scales $\hlam < \hld$ scales as
 \beq
 \delta \hat{z} \sim \hld^{1/3} \left(\hlam/\hld\right)^{\zot},\label{eq:inbetween}
 \eeq
where $\zeta_2^\perp = 1/2$ in the MS17 model (and also in the original B06 theory) and $\zeta_2^\perp = 0.52$ in the CSM15 model. 

These structures will eventually, in turn, be disrupted at a secondary disruption scale, have their amplitude dynamically adjusted to keep the energy flux constant and their perpendicular anisotropy removed, engendering another "mini-cascade", and so on. Therefore, what we have so far called $\hld$ is only the first of many subsequent disruption scales -- and so from now on, we will call this first disruption scale $\hldd$. We can therefore identify two distinct subranges of MHD turbulence:
\begin{align}
\hlam > \hlam_{{\rm D},1}, \, &\text{"free alignment range"},\nonumber\\
\hlam < \hlam_{{\rm D},1}, \, &\text{"disruption range"}.\nonumber
\end{align}
The two subranges are shown in Figure \ref{fig:cascade}. We now proceed to discuss the sequence of disruptions (Section \ref{sec:recurse}), the dissipation scale $\hlam_{\eta,\infty}$ (Section \ref{sec:etainf}), and the spectral index in the disruption range (Section \ref{sec:spectrum}).

\subsection{Recursive disruption}\label{sec:recurse}
\begin{figure}
\vskipfig
\includegraphics[width=\linewidth]{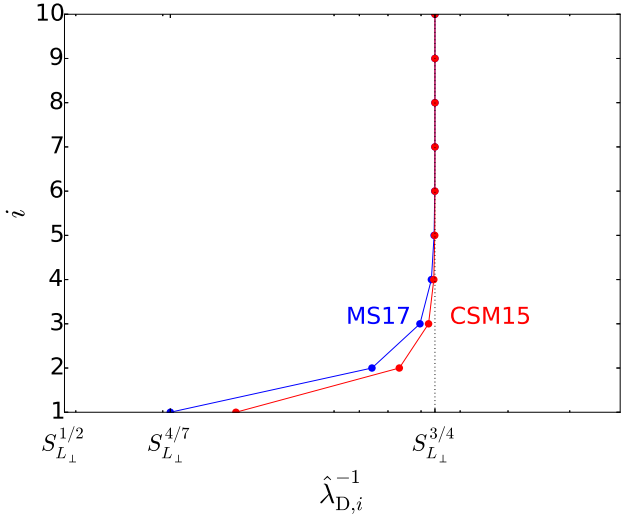}
\vskipfig
\caption{The dependence of $\hldp$ on $i$ [Eq.~(\ref{eq:hldp2})] is shown for the MS17 model (blue) and for the CSM15 model (red). The Kolmogorov scale given by Eq.~(\ref{eq:hletainf}) is shown as a black dotted line. \label{fig:hldp}}
\vskipfig
\end{figure}

The series of consecutive disruptions can be understood as follows. After the $(i-1)$st disruption, the turbulence behaves as though there is an $i$th cascade, with "outer-scale" values of the turbulent variables given by the values at the $(i-1)$st disruption scale, $\hlam_{{\rm D},i-1}$. The cascade has the same scalings as the original cascade, but with the replacements
\begin{align}
L_\perp \rightarrow \lambda_{{\rm D},i-1},\quad
\overline{\delta z} \rightarrow \delta z_{i-1,-}\sim \hldpm^{1/3}\overline{\delta z} ,\label{eq:repl}
\end{align}
in all places where either of these variables appear (including normalizations).
%In between the disruptions, the amplitude in the $i$th cascade scales as [cf. Eq~(\ref{eq:inbetween})]
%\beq
%\delta \hat{z} \sim \hldpm^{1/3} \left(\hlam / \hldpm \right)^{\zot}.
%\eeq
%and so one can see that the amplitudes \emph{just below} the disruption scales $\delta z_{i,-}$ follow the scaling one expects from the \citealt{gs95} (GS95) model, but \emph{between} disruptions every fluctuation is dynamically aligning and follows the \citet{boldyrev} scaling. Since in practice the disruption happens to progressively more fluctuations over a range of scales, the structure functions and spectrum are likely to in practice have scalings close to the GS95 prediction, i.e. a spectral index of $-5/3$. Effectively, the disruption process puts a physical lower limit on the alignment of the fluctuations, and so overall the turbulence in the disruption range is forced to have the GS95 scalings that one would expect without scale-dependent dynamic alignment.
Using the rule (\ref{eq:repl}) in Eqs.~(\ref{eq:ndms}), (\ref{eq:ndcsm}) for $\hld$ leads to the recursive relation
\begin{align}
\hldp &\sim \hldpm^{1+4\fsd/3}\SL^{\fsd}.\label{eq:hldp}
\end{align}
This scale has been normalized by $L_\perp$ after performing the replacement procedure (\ref{eq:repl}). The exponent $\fsd$ depends on the choice of turbulence model, and is the exponent in Eq.~(\ref{eq:ldmeq2}):
\beq
\fsd = \begin{cases}
-4/7, \, &\text{MS17 \& B06 models}, \\
%-0.45\left(1-0.52{\zeta_2^\perp}\right)^{-1}
-0.62, \, &\text{CSM15 model}.
\end{cases}
\eeq
The recursion relation (\ref{eq:hldp}) may be written as 
\begin{align}
%\hldp &\sim \hlam_{{\rm D},1}^{\left(1+4\fsd/3\right)^{i-1}}\SL^{\quad\displaystyle\fsd \sum_{j=0}^{i-2} \left[1+4\fsd/3\right]^j}.\label{eq:hldp2}
\hldp &\sim \SL^{\,\fsd\sum_{j=0}^{i-1} \left[1+4\fsd/3\right]^j}.\label{eq:hldp2}
\end{align}
As $i\to\infty$, we have
\beq
\hlam_{{\rm D},\infty} \sim \SL^{-3/4}.\label{eq:hldinf}
\eeq
Figure \ref{fig:hldp} shows the scale $\hlam_{{\rm D},i}$, for $i=1,2,3,...,10$. Obviously, as $i$ increases, the successive disruptions become ever closer to each other in scale, and so the disruption scale quickly approaches the asymptotic value (\ref{eq:hldinf}).

\subsection{Final dissipative cutoff scale}\label{sec:etainf}
Similarly to Eq.~(\ref{eq:hldp}), using the rule (\ref{eq:repl}) in Eqs.~(\ref{eq:hletams}), (\ref{eq:hletacsm}) for $\hlam_{\eta}$ leads to the recursive relation
\begin{align}
\hletap &\sim \hldpm^{1+4\feta/3}\SL^{\feta},\label{eq:hletap}
\end{align}
where the exponent $\feta$ is given by the exponents of Eqs.~(\ref{eq:hletams}) or (\ref{eq:hletacsm}): 
\beq
\feta = \begin{cases}
-2/3, \, &\text{MS17 \& B06 models}, \\
%-\left(1.60-0.32{\zeta_2}\right)^{-1}
-0.70, \, &\text{CSM15 model, $m=2$}.
\end{cases}
\eeq
The limit of $\hletap$ as $i\to\infty$ is also given by (\ref{eq:hldinf}), so
\beq
\hlam_{{\eta},\infty} \sim \hlam_{{\rm D},\infty} \sim \SL^{-3/4}.\label{eq:hletainf}
\eeq
Since $\fsd > \feta$ for both models, $\hldp>\hletap$ for all $i<\infty$, and $\hlam_{\eta,\infty}$ may be considered the final dissipation scale for the cascade. This scale is the same as the \citet{k41} scale that one expects as the dissipation scale in the GS95 model, i.e., for MHD turbulence without scale-dependent dynamic alignment. This reflects the fact that there is a lower limit on alignment imposed by the disruption process. This dissipation scale is the key testable prediction of our model for the disruption range. Encouragingly, \citet{Beresnyak14} found that in numerical simulations of RMHD turbulence, the dissipation scale was very close to the scale $\hlam_{\eta,\infty}$.

\subsection{Coarse-grained spectrum}\label{sec:spectrum}
We will now proceed to estimate the effective spectral index of the turbulent fluctuations in the disruption range. %Since the recursive disruption process we have described involves "jumps" in the amplitude at each disruption, the spectrum in the disruption range is not a true power law. Nevertheless, 
Namely, we will examine the amplitudes just above and just below the disruption scales to bound the effective scaling exponent in the disruption range. 

The "lower amplitude", just below the $i$th disruption, scales as [cf. Eq.~(\ref{eq:under})]
\beq
\delta \hat{z}_{i,-} \sim \hldp^{1/3}.\label{eq:lower}
\eeq
As $i\to\infty$, $\delta\hat{z}_{\infty,-}\sim\SL^{-1/4}$. These lower amplitudes, defined only on the coarse-grained set of scales $\hldp$, define the lower envelope of the second-order structure function (or spectrum).
\begin{figure}
\vskipfig
\includegraphics[width=\linewidth]{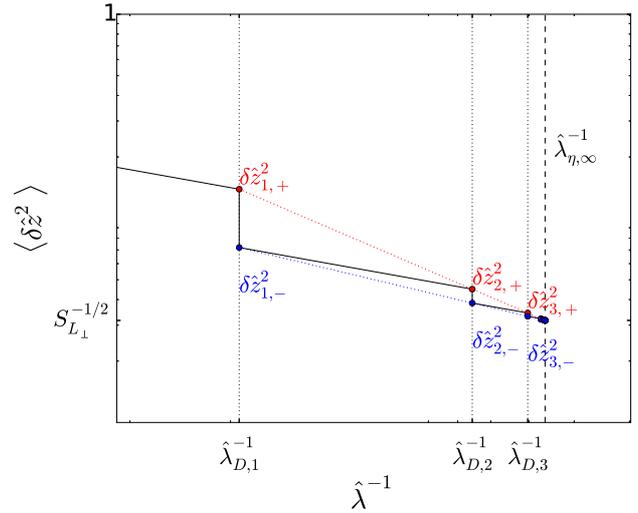}
\vskipfig
\caption{Schematic showing the idealized form of the second-order structure function in the disruption range (black solid line). Also shown are the upper amplitudes (\ref{eq:upper}) as red points, lower amplitudes (\ref{eq:lower}) as blue points, and the upper and lower envelopes as red and blue dotted lines respectively. The first three disruption scales are marked with vertical dotted lines, and the final dissipation scale $\hlam_{\eta,\infty}$ (\ref{eq:hletainf}) is marked with a vertical dashed line. We stress that the true structure function will be continuous and somewhere between the upper and lower envelopes.\label{fig:sf}}
\vskipfig
\end{figure}

The "upper amplitude", just above the $i$th disruption, scales as [cf. Eq.~(\ref{eq:inbetween})]
\beq
\delta \hat{z}_{i,+} \sim \hldpm^{1/3} \left(\hldp / \hldpm \right)^{\zot}.
\eeq
Using the recursion relation Eq.~(\ref{eq:hldp}), this may be written as
\beq
\delta \hat{z}_{i,+} %\sim \hldp^{\frac{1}{3}(1+\fsd)(1+4\fsd/3)^{-1}} \SL^{\fsd\left[\frac{\zeta_2^\perp}{2} - \frac{1}{3} (1+\fsd)(1+4\fsd/3)^{-1}\right]}.
\sim \hldp^{(1/3 -\zot)(1+4\fsd/3)^{-1} + \zot} \SL^{-\fsd(1/3 -\zot)(1+4\fsd/3)^{-1}}.\label{eq:upper}
\eeq
In the MS17/B06 model, this is
\beq
\delta \hat{z}_{i,+}^{\rm MS} \sim \hldp^{3/5}\SL^{1/5},\label{eq:msupper}
\eeq
while in the CSM15 model, 
\beq
\delta \hat{z}_{i,+}^{\rm CSM}\sim \hldp^{0.68}\SL^{0.26}.\label{eq:csmupper}
\eeq
As $i\to\infty$, $\delta \hat{z}_{\infty,+}\sim\SL^{-1/4}$ for both models, the same as the lower amplitudes. The upper amplitudes, defined on the coarse-grained set of points $\hldp$, determine the upper envelope of the second-order structure function. Between disruptions, the fluctuations dynamically align and have the corresponding $\delta\hat{z}\propto\hlam^{\zot}$ scaling. A schematic for the idealized second-order structure function is shown in Figure~\ref{fig:sf}. It consists of segments with the scaling $\zeta_2^\perp$, joined by discontinuous jumps at each disruption scale $\hldp$. In reality, the true structure function will be continuous and lie between the upper and lower envelopes.

The effective scaling of the second-order structure function is therefore bounded above by $\hlam^{6/5}$ [MS17/B06, Eq.~(\ref{eq:msupper})]  or $\hlam^{1.3}$ [CSM15, Eq.~(\ref{eq:csmupper})], and below by $\hlam^{2/3}$ [Eq.~(\ref{eq:lower})]. Using the usual correspondence between the second-order structure function and the spectrum, we expect the effective spectral index in the disruption range to be between $-5/3$ and $-2.3$ (CSM15) or $-11/5$ (MS17/B06) in this range\footnote{It might be worth mentioning in this context the results of \citet{beresnyak2013reconnection} and \citet{kowal2016}, who observed in 3D numerical simulations that reconnecting sheets generate turbulence that seems to agree with the GS95 scalings, and also the results of \citet{huang2016}, who performed a different simulation of reconnection-driven turbulence and found turbulence with a perpendicular spectral index of $-2.1$ to $-2.3$.}. This is significantly steeper than the $-3/2$ in the free alignment range, despite the fact that between disruptions, there is scale-dependent alignment of fluctuations in a similar way to the primary cascade. However, to measure such a scaling unambiguously, one would likely need extremely high $\SL$, high enough to have a good scale separation between $\hldd\sim \SL^{-4/7}\sim\SL^{-0.6}$ and $\hlam_{\eta,\infty}\sim\SL^{-3/4}$. Thus, to test our model for the disruption range, it is potentially more productive to determine the scaling of the dissipation scale $\hlam_{\eta,\infty}$, comparing it to the $\SL^{-3/4}$ scaling given in Eq.~(\ref{eq:hletainf}).

It is perhaps worth commenting on how one might expect the scaling of the traditional alignment angles based on ratios of structure functions involving angles between different RMHD fields \citep{mcatbolalign,bl06} to change in the disruption range. Because there is a physical lower limit to the alignment angle of turbulent structures in this range, these alignment measures will likely have a shallower scaling exponent at scales below $\hldd$. Shallower scaling exponents for these measures were indeed observed at the smallest scales in the numerical simulations of both \citet{pmbc,Perez14} and \citet{beresnyak}.

\section{Discussion}
The dynamic-alignment models of strong Alfv\'enic turbulence due to \cite{boldyrev}, \cite{Chandran14} and \cite{ms16} all predict that, as turbulent structures cascade to smaller scales, the vector fluctuations within them progressively align, and the structures become progressively more sheet-like and anisotropic within the perpendicular plane.
In this paper, inspired by the recent work on the disruption of forming current sheets by \cite{uzdensky2016}, we have found that these sheet-like structures are destroyed by reconnection below a certain scale $\hld$. This disruption process occurs in two stages: linear growth of a tearing instability with multiple islands, and then collapse of the $X$-points between these islands into thin current sheets, which reconnect until the original structure has been destroyed. This means that the linear growth rate must be large compared to the cascade rate of the turbulence in order for the structures to be disrupted. To estimate the timescales involved, we have used scalings from the turbulence models of \cite{ms16} and \cite{Chandran14}. Qualitatively, these models give similar results, although quantitatively the predicted scalings are slightly different.

We find that there is a critical scale $\hld\sim \SL^{-0.6}$, below which the turbulent structures are disrupted (see Section \ref{sec:transition}). This means that the turbulence theories that rely on dynamic alignment can only be expected to give accurate predictions at scales above $\hld$. At $\hld$, the turbulent cascade is effectively reset to unaligned structures, which can now cascade to smaller scales and again become progressively more sheet-like and aligned. We show that they are recursively disrupted at a sequence of smaller scales $\hldp$, with $i=2,...,\infty$ (see Section \ref{sec:recurse}). We place bounds on the effective spectral index in the "disruption range" below $\hld$, and show that the effective spectral index is between $-5/3$ and $-2.3$, significantly steeper than the approximately $-3/2$ spectral index above $\hld$ (see Section \ref{sec:spectrum}). However, a very large $\SL$ is needed to detect a reliable power law in this range. 

The disruptions get progressively closer to each other in scale as $i$ increases, and in the limit $i\to\infty$ 
the turbulent fluctuations reach a final dissipation scale $\hlam_{\eta,\infty}\sim \SL^{-3/4}$ (see Section \ref{sec:etainf}). This is a smaller scale than the dissipation scale predicted by the dynamic-alignment theories \citep{boldyrev,Chandran14,ms16}, and is identical to the \citet{k41} scale that one would expect for turbulence with a $-5/3$ spectrum (i.e., in the absence of dynamic alignment). This is despite the fact that the spectral index above $\hlam_{{\rm D}}$ in our model is approximately $-3/2$ typical of the dynamic-alignment theories, and that between disruptions, there is scale-dependent alignment: effectively, the disruption process imposes a physical lower limit on the alignment angle.
Thus, our argument that sheet-like structures are disrupted by reconnection below $\hld$ might explain the discrepancy between the measured $-3/2$ spectrum in numerical simulations \citep{Perez14}, and the seemingly opposing evidence that the dependence of the dissipation range on viscosity or resistivity\footnote{All relevant simulations were done with equal viscosity and resistivity.} is much better described by the Goldreich-Sridhar/Kolmogorov scaling \citep{Beresnyak14}. Effectively, both sets of measurements are correct, but neither tells the "full story": at large scales, dynamic alignment does occur, but at sufficiently small scales, the sheet-like structures become unstable, which limits the alignment, steepens the spectrum and forces the dissipation scale to have the Kolmogorov scaling. The scaling of $\hlam_{\eta,\infty}$ is the key prediction of our model that is testable in currently feasible numerical simulations.

There are many improvements possible to the simple model of the disruption process and of the "disruption range" that we have proposed here. First, we have neglected the effects of shear and viscosity on the stability of current layers \citep{chen1990res,chen1990visc}. Secondly, our conjectures about the turbulence below $\hlam_{{\rm D}}$ are rather simple: we completely ignore the intermittency in this range, and do not take into account anything about the specific nature of the "flux-rope-like" structures produced by the tearing instabilities, apart from conjecturing a limit on their anisotropy in the perpendicular plane. Thirdly, we ignore any potential dissipation by the reconnection process; this may steepen the spectral index in the disruption range. Finally, in many situations (including the solar wind), kinetic scales will intervene at some point in the collapse process, significantly altering the dynamics. Nevertheless, we expect the idea that the sheet-like structures produced by dynamically aligning turbulence will eventually reconnect and destroy themselves is robust, even in kinetic systems, and provides an interesting link between inertial-range intermittent turbulent structures and magnetic reconnection.

\section*{Acknowledgements} We thank A. Beresnyak and N. Loureiro for useful conversations, and the referee, D. Uzdensky, for helping us improve the manuscript. While this manuscript was in an advanced stage of preparation, we became aware that a similar calculation of the disruption scale of sheet-like structures in Alfv\'enic turbulence [our Eq.~(\ref{eq:ldmeq2})] was concurrently being done by N. Loureiro and S. Boldyrev \citep{loureiroboldyrev}. We are grateful to N. Loureiro for alerting us to this work. The work of A.M. and B.D.G.C. was supported by NASA grant NNX15AI80G and NSF grants PHY-1500041 and AGS-1258998. A.M. was additionally supported by NSF grant AGS-1624501. The work of A.A.S. was supported in part by grants from UK  STFC and EPSRC. A.M. and A.A.S. would like to acknowledge the hospitality of the Wolfgang Pauli Institute, Vienna, where the idea behind this work was first conceived.

\bibliographystyle{mnras}
\bibliography{mainbib2}
\end{document}